\documentclass[11pt]{article}
\usepackage{clay}
\usepackage{hyperref}
\hypersetup{colorlinks=true}
\hypersetup{linkcolor=black}
\hypersetup{citecolor=black}
\hypersetup{urlcolor=black}

\usepackage{tikz}
\usetikzlibrary{arrows,decorations.markings,cd}
\tikzset{
->-/.style={decoration={markings, mark=at position .5 with {\arrow[scale=1.5]{stealth}}}, postaction={decorate}}
}
\usepackage{epsfig}
\usepackage{amssymb}
\usepackage{amsfonts}
\usepackage{amsbsy}
\usepackage{amsmath}
\usepackage{dsfont}
\usepackage{bbm}
\usepackage{upgreek}
\usepackage{amscd}
\usepackage{graphicx}
\usepackage{mathrsfs}
\usepackage{amsmath,amsthm}
\usepackage{slashed}
\usepackage{dsfont}
\usepackage{tikz}
\usepackage[utf8]{inputenc}
\usepackage{tikz-feynman}\tikzfeynmanset{compat=1.0.0}
\numberwithin{equation}{section}

\usetikzlibrary{positioning}
\usetikzlibrary{shapes.geometric, arrows}
\tikzstyle{rect} = [rectangle,rounded corners,minimum width=3cm, minimum height=1cm, text centered, draw=black]
\tikzstyle{arrow} = [thick,->,>=stealth]
\usepackage{stackrel}
\tikzstyle{oct} = [regular polygon,regular polygon sides=8, draw,
    text width=2em, text centered]
\tikzstyle{line} = [draw, -latex']

\tikzstyle{hex} = [regular polygon,regular polygon sides=6, draw,
    text width=2em, text centered]
\tikzstyle{hexs}= [regular polygon,regular polygon sides=5, draw,
    text width=1.7em, text centered]

\makeatletter
\DeclareFontFamily{OMX}{MnSymbolE}{}
\DeclareSymbolFont{MnLargeSymbols}{OMX}{MnSymbolE}{m}{n}
\SetSymbolFont{MnLargeSymbols}{bold}{OMX}{MnSymbolE}{b}{n}
\DeclareFontShape{OMX}{MnSymbolE}{m}{n}{
    <-6>  MnSymbolE5
   <6-7>  MnSymbolE6
   <7-8>  MnSymbolE7
   <8-9>  MnSymbolE8
   <9-10> MnSymbolE9
  <10-12> MnSymbolE10
  <12->   MnSymbolE12
}{}
\DeclareFontShape{OMX}{MnSymbolE}{b}{n}{
    <-6>  MnSymbolE-Bold5
   <6-7>  MnSymbolE-Bold6
   <7-8>  MnSymbolE-Bold7
   <8-9>  MnSymbolE-Bold8
   <9-10> MnSymbolE-Bold9
  <10-12> MnSymbolE-Bold10
  <12->   MnSymbolE-Bold12
}{}

\let\llangle\@undefined
\let\rrangle\@undefined
\DeclareMathDelimiter{\llangle}{\mathopen}%
                     {MnLargeSymbols}{'164}{MnLargeSymbols}{'164}
\DeclareMathDelimiter{\rrangle}{\mathclose}%
                     {MnLargeSymbols}{'171}{MnLargeSymbols}{'171}
\makeatother

\def\be{ \begin{equation} }
\def\ee{ \end{equation}}

\newcommand{\eq}[1]{\begin{align}\begin{split}#1\end{split}\end{align}}

\def\exp{{\rm exp}}

\def\log{{\rm log}}

\def\one{{\hbox{ 1\kern-.8mm l}}}
\def\CO {{\cal O}}

\def\CO {{\cal O}}

\newcommand{\bfg}{{\bf g}}
\newcommand{\bfh}{{\bf h}}
\newcommand{\bfk}{{\bf k}}

\def\IZ{{\mathbb{Z}}}

\def\fe{\mathfrak{e}}

\def\rmk#1{\bigskip\noindent{\bf Remark} }
\def\cnj#1{\bigskip\noindent{\bf Conjecture:} }
\DeclareMathAlphabet{\mathpzc}{OT1}{pzc}{m}{it}

\def\Tr{ \, \textrm{Tr} \, }

\def\ie{\begin{equation}\begin{aligned}}
\def\fe{\end{aligned}\end{equation}}

\begin{document}

\date{November, 2020}

\institution{UC}{\centerline{Kadanoff Center for Theoretical Physics \& Enrico Fermi Institute, University of Chicago
}}

\title{Axions, Higher-Groups, and Emergent Symmetry }

\authors{T. Daniel Brennan\footnote{e-mail: {\tt tdbrennan@uchicago.edu}} and Clay C\'{o}rdova\footnote{e-mail: {\tt clayc@uchicago.edu}}}
\abstract{ Axions, periodic scalar fields coupled to gauge fields through the instanton density, have a rich variety of higher-form global symmetries.  These include a two-form global symmetry, which measures the charge of axion strings.  As we review, these symmetries typically combine into a higher-group, a kind of non-abelian structure where symmetries that act on operators of different dimensions, such as points, lines, and strings, are mixed.  We use this structure to derive model independent constraints on renormalization group flows that realize theories of axions at long distances. These give universal inequalities on the energy scales where various infrared symmetries emerge. For example, we show that in any UV completion of axion-Yang-Mills, the energy scale at which axion strings can decay is always larger than the mass scale of charged particles.

 }

\maketitle

\setcounter{tocdepth}{3}
	
\tableofcontents

\section{Introduction}
\label{seci}

Axions are are ubiquitous in models of beyond the standard model physics and string theory, and have been a central focus of a variety of experiments. (See e.g.\ \cite{Kim:2008hd, Marsh:2015xka, Svrcek:2006yi} and references therein.)  One of the key phenomena in the physics of axions is the existence of string-like excitations.  Let $a$ denote the periodic axion field with periodicity $2\pi f$ where $f$ is the axion decay constant.  Around a string of charge $n$ the axion field winds as
\begin{equation}
\oint da =2\pi n f~. \label{windingstring}
\end{equation}
In a low energy model of axions, the strings are non-dynamical defects with infinitesimal thickness.  At higher energies these theories have a rich array of new degrees of freedom.  In particular, the axion is often revealed to be the angular part of a complex field $\varphi$ which is liberated in the ultraviolet. Relatedly, at higher energies the strings are brought to life as dynamical excitations of the field theory.  For instance at the scale set by the string tension $T^{1/2}_{\mathrm{string}}$, strings may nucleate from the vacuum. 

Thus, quantum field theories with axions in their long distance physics are intrinsically equipped with an approximate energy scale $E_{\mathrm{string}}$ above which the string quantum number is no longer conserved. This scale is less than that set by the string tension, but in perturbative models of axions these scales are often far separated.  One of our goals in the following will be to derive model independent constraints on the scale $E_{\mathrm{string}}$ applicable to any UV completion of axion physics.

It is fruitful to describe aspects of axion strings in the language of generalized global symmetries \cite{Gaiotto:2014kfa}.  In this terminology, the periodic scalar field $a$ gives rise to a conserved current of a generalized two-form global symmetry $U(1)^{(2)}$:\footnote{A continuous $p$-form global symmetry has a conserved current which is a $(p+1)$-form.  The charged objects are extended operators of dimension $p$. Familiar ordinary global symmetries are described by the special case when $p=0$. }
\begin{equation}
J_{\mu\nu\rho}\sim\varepsilon_{\mu\nu\rho \sigma}\partial^{\sigma}a~, \label{J2def}
\end{equation}
which is tautologically conserved. The axion strings described above carry the charge associated to this current and equation \eqref{windingstring} expresses the Ward identity of a current in the presence of a charged object. The current defined by \eqref{J2def} should be distinguished from the frequently discussed current $J_{\alpha}=\partial_{\alpha}a$ for the shift symmetry of the axion. The latter is spontaneously broken by the choice of vacuum in the theory (with $a$ the associated Goldstone mode).  Meanwhile the higher-dimensional analog of the Coleman-Mermin-Wagner theorem \cite{Gaiotto:2014kfa, Lake:2018dqm} implies that the symmetry associated to $J_{\mu\nu\rho}$ is never spontaneously broken at long distances.\footnote{Recall from \cite{Coleman:1973ci}, that in two spacetime dimensions a continuous global symmetry cannot be spontaneously broken because the associated Goldstone boson would necessarily have an unphysical logarithmic two-point function. The statement above is the direct dimensional uplift to four spacetime dimensions.}

The hallmark of an axion is its coupling to gauge fields. Consider for instance axion-Yang-Mills with gauge group $SU(N)$ and action:
\begin{equation}\label{aymactintro}
S=\frac{1}{2}\int da \wedge *da +\frac{1}{g^{2}}\int \mathrm{Tr}(F\wedge *F)-\frac{i}{8\pi^{2}f}\int a \mathrm{Tr}(F\wedge F)~.
\end{equation}
In addition to the two-form symmetry $U(1)^{(2)}$ described above, this theory also has a discrete one-form symmetry $\mathbb{Z}_{N}^{(1)}$.  The objects charged under this discrete symmetry are Wilson lines, and the conserved quantum number is the charge under the $\mathbb{Z}_{N}$ center of the gauge group.  This is preserved by interactions because, in the absence of additional matter, the dynamical gluons can only screen adjoints and other representations of vanishing charge. This one-form symmetry is also intimately connected to confinement: at long distances it is preserved or broken precisely when the gauge theory is in a confined or deconfined phase respectively \cite{Gaiotto:2014kfa}.

Thus, the presence of the one-form symmetry $\mathbb{Z}_{N}^{(1)}$, endows UV completions of axion-Yang-Mills with another important energy scale, $E_{\mathrm{screen}},$ above which the one-form symmetry is broken.  Physically, this is the energy scale where matter fields charged under $SU(N)$ appear and hence can screen general Wilson lines. Below in section \ref{secemerge} we will show that, independent of the details of the ultraviolet physics, there is an inequality on these symmetry breaking scales $E_{\mathrm{string}}$ and $E_{\mathrm{screen}}$:
\begin{equation}
E_{\mathrm{screen}}\lsim E_{\mathrm{string}}~. \label{result1}
\end{equation}
Of course, since they are defined by emergent symmetries, the energy scales above are in general only approximately meaningful.  Far below $E_{i}$ the corresponding quantum number is preserved up to small effects, while far above $E_{i}$ the symmetry is badly broken.  We thus interpret \eqref{result1} as forbidding a parametric separation of scales violating the inequality.  In other words, $E_{\mathrm{screen}}\gg E_{\mathrm{string}}$ is not possible in any renormalization group trajectory, which flows at long distances to axion-Yang-Mills described by the action \eqref{aymactintro}.

To argue for the inequality \eqref{result1} and other similar bounds, we first study in detail the symmetry structure of theories of axions in section \ref{sec:2}.  As we describe, the various symmetries of these models do not factorize into a simple product of the symmetries of each form degree.  Instead, a more precise analysis shows that they form a mixed structure known as a ``higher group" global symmetry, which is somewhat analogous to a non-abelian symmetry group composed of symmetries of different degrees.  

Higher-group global symmetry is an analog of the Green-Schwarz mechanism \cite{Green:1984sg} for global symmetries (as opposed to its more common application when all gauge fields are dynamical) \cite{Kapustin:2013uxa, Kapustin:2014zva}.  This symmetry structure frequently appears in gauge theories and in topological quantum field theories when symmetries of differing form degrees are present \cite{Tachikawa:2017gyf, Cordova:2018cvg, Benini:2018reh}.  In the context of models of axions, the presence of these interesting symmetry structures has been noted for non-abelian gauge groups in \cite{Seiberg:2018ntt, Cordova:2019uob}, and recently explored in abelian gauge theories in \cite{Hidaka:2020iaz, Hidaka:2020izy}, and our presentation in section \ref{sec:2} draws closely from these references.  Our analysis leading to inequalities on energy scales using higher-group symmetry follows closely the logic of \cite{Cordova:2018cvg}, and more broadly the spirit of \cite{Cordova:2020tij}.

One straightforward way to make the presence of this higher-group transparent is by coupling the theory to background gauge fields for the global symmetries.  For example, in axion Yang-Mills the $U(1)^{(2)}$ two-form symmetry that measures the charges of strings, has a natural source which is a three-form gauge field $A^{(3)}$ with gauge-invariant field strength $G^{(4)}$, while the $\mathbb{Z}_{N}^{(1)}$ one-form symmetry that measures the charges of Wilson lines is sourced by a two-form gauge field $B^{(2)}$ which has periods that are $N$-th roots of unity.  The higher-group structure derived below then implies that the field strength $G^{(4)}$ is:
\begin{equation}\label{g4intro}
G^{(4)}=dA^{(3)}+\frac{N(N-1)}{4\pi} B^{(2)}\wedge B^{(2)}~.
\end{equation}
Notice in particular that the presence of the one-form symmetry background induces non-trivial (and in general fractional) $G^{(4)}$ flux.  This mixing between sources for global symmetries is the signature of higher-group global symmetry.

In section \ref{secemerge} we derive the inequality \eqref{result1} using the flux formula \eqref{g4intro}. We then investigate the bound in the context of weakly-coupled KSVZ-type models.  In such theories axions and the symmetry structure above emerge at low-energies after breaking an anomalous Peccei-Quinn symmetry \cite{Kim:1979if,Shifman:1979if,Peccei:1977hh,Peccei:1977ur}. As we show, the inequality \eqref{result1} implies constraints on ultraviolet coupling constants which are indeed true in theories that flow to axion-Yang-Mills in the IR.

In section \ref{secemerge} we also discuss constraints on axion-Maxwell theories.  We consider the action:
\begin{equation}
S=\frac{1}{2}\int da \wedge *da+\frac{1}{2g^{2}}\int F\wedge *F-\frac{i K}{8\pi^{2}f}\int a F\wedge F~,
\end{equation}
where above $K\in \mathbb{Z}$ is an integral coupling constant.  In particular, for $|K|>1$ we find a non-trivial higher-group global symmetry and resulting constraints on emergent symmetry scales.  The symmetries involved are of two types.  First, there is the continuous $U(1)^{(2)}$ string symmetry of the axion and the $U(1)_{m}^{(1)}$ magnetic symmetry of the photon that measures the magnetic charges of 't Hooft lines.  There is also a discrete $\mathbb{Z}_{K}^{(0)}$ shift-symmetry of the axion and a discrete $\mathbb{Z}_{K}^{(1)}$ one-form symmetry of the photon that measures the charges of Wilson lines modulo $K$.  In addition to results similar to \eqref{result1} we also derive the general constraint
 \begin{equation}\label{result2}
\min\{E_{\mathrm{shift}}, ~E_{\mathrm{screen}}\} \lsim E_{\mathrm{magnetic}}~,
\end{equation}
where above, $E_{\mathrm{shift}}$ is the scale where the shift symmetry of the axion is violated, $E_{\mathrm{screen}}$ is the scale at which Wilson lines can be screened by charged matter, and $E_{\mathrm{magnetic}}$ is the scale of emergence of the $U(1)$ gauge field defining the photon, above which $*F$, ceases to define a conserved current.  We verify \eqref{result2} in a simple example and find that the implied constraints on coupling constants are indeed necessary for self-consistency of our analysis.

Finally, we also discuss analogs of \eqref{result1} and \eqref{result2} in theories with axions and charged matter fields that have higher-group global symmetry and hence the ensuing consequences.

\section{Symmetry Structure in Models of Axions}

\label{sec:2}

In this section we review the basic definitions of higher-form global symmetry following \cite{Gaiotto:2014kfa}, and higher-group global symmetry following \cite{Cordova:2018cvg, Benini:2018reh}. We then present the symmetries and couplings to background gauge fields in models of axions. In the case of axions coupled to abelian gauge fields these symmetries have been previously studied in \cite{Hidaka:2020iaz, Hidaka:2020izy}.  In the case of axions coupled to non-abelian gauge fields, our analysis follows from results in \cite{Seiberg:2018ntt, Cordova:2019uob}.

\subsection{Higher-Form Symmetry}

In general, a $p$-form global symmetry is a symmetry that acts on extended operators of dimension $p$ \cite{Gaiotto:2014kfa}. Thus, ordinary global symmetries, which act on point operators, correspond to the case $p=0$.  Meanwhile one-form symmetries act on line operators, and two-form symmetries act on surface operators.  We will encounter all of these symmetries below.  Higher-form symmetries form a group (necessarily abelian if $p>0$) that we denote $G^{(p)}$. In the simplest case of continuous $G^{(p)}$, there is a conserved current with $p+1$ antisymmetric indices:
\begin{equation}
\partial^{\mu^{1}}J_{\mu_{1} \cdots \mu_{p+1}}=0~.
\end{equation}
It is frequently convenient to work with the hodge dual $*J$ which is therefore closed, $d*J_{p+1}=0.$

As usual when discussing the implications of symmetry it is fruitful to introduce background gauge fields that act as sources for the conserved currents. For a $p$-form global symmetry the appropriate background is a $(p+1)$-form gauge field $A^{(p+1)}$. In the simplest case of a continuous $G^{(p)}$ symmetry, the background gauge field couples in a standard way to the conserved current:
 \begin{equation}
 S \supset i\int A^{(p+1)}\wedge \ast J_{p+1}~. \label{sourcecurr}
 \end{equation}
Current conservation means that (up to possible 't Hooft anomalies) the theory is invariant under background gauge transformations of the source:
\begin{equation}\label{vgauge}
A^{(p+1)}\longmapsto A^{(p+1)}+d\Lambda^{(p)}~,
\end{equation}
where $\Lambda^{(p)}$ is locally a $p$-form gauge parameter.  The normalization of the coupling \eqref{sourcecurr} between the source and the current is fixed by charge quantization.  In our conventions $G^{(p)}$ charges, which are measured on any closed surface $\Sigma_{d-p-1}$ of dimension $(d-p-1)$, are integrally quantized:
\begin{equation}
\int_{\Sigma_{d-p-1}}*J_{p+1} \in \mathbb{Z}~.
\end{equation}
In particular, this means that the fluxes and gauge parameters of the background fields are also quantized:
\begin{equation}
\int_{\Sigma_{p+2}} \frac{dA^{(p+1)}}{2\pi}\in \mathbb{Z}~, \hspace{.5in}\int_{\Sigma_{p+1}} \frac{d\Lambda^{(p)}}{2\pi}\in \mathbb{Z}~.
\end{equation}

Alternatively, we can also describe symmetry through the symmetry group operators.  For an element $g\in G^{(p)},$ the associated operator is denoted $U_{g}(\Sigma_{d-p-1})$ and is defined on any closed $\Sigma_{d-p-1}$. In the case of a continuous symmetry these are given by exponentiated integrals of the current
\begin{equation}\label{symops}
U_g(\Sigma_{d-p-1})=e^{i \lambda \int_{\Sigma_{d-p-1}} *J_{p+1}}~,
\end{equation}
where $g=e^{i\lambda}$. The primary advantage of the symmetry group operators is that the continue to exist even when the symmetry is discrete.  Mathematically these symmetry defect operators can be thought of as Poincar\'{e} dual to the background fields.  They describe flat background gauge fields, i.e.\ those with vanishing field strength.

\subsubsection{Higher-Group Global Symmetry}
\label{hgreview}

Higher-group global symmetry is a natural possibility in theories with symmetries of differing form degrees.  As in the discussion above, there are various related point of view in terms of currents, symmetry group operators, and background fields.

Perhaps the simplest context where this mixing occurs is in situations where all symmetries in question are continuous and so we can reduce to statements about the conserved currents.  In this case we can find contact terms in the operator product expansion of the generic form:
\begin{equation}\label{hgrpcur}
d*J_{p_{1}+1}(x) J_{p_{2}+1}(y)\sim \partial^{k}\delta^{(d)}(x-y)J_{p_{3}+1}(x)~,
\end{equation}
where $\partial^{k}$ is a $k$-th derivative and $p_{3}=p_{1}+p_{2}+k$.  Note in particular that, since $k\geq 0,$ the fusion always produces symmetries of form degree at least $p_{1}+p_{2}$ on the right-hand-side.  Such fusion laws are similar in spirit to the more familiar non-abelian current algebra of ordinary (0-form) global symmetries.  In that context, the currents are conserved at separated points, but at coincident points one encounters a contact term in the divergence controlled by the Lie algebra structure constants $f^{abc}$.  Equation \eqref{hgrpcur} is similar except that now the fusion involves currents of differing form degrees and there are derivatives on the delta function.  For instance, an example of this structure, explored in detail in \cite{Cordova:2018cvg}, occurs in multiflavor massless QED, where the zero-form flavor symmetry forms a higher-group with the one-form magnetic symmetry with current $J_{2}=*F$.  Other examples are described in detail in \cite{Cordova:2020tij}.

In the applications below, we will often be interested in higher-groups where the symmetries involved are discrete.  In this case we read off the structure from the symmetry group operators $U_{g}(\Sigma)$ described above following \cite{Benini:2018reh}.  To begin with, we recall from \cite{Gaiotto:2014kfa} that the group multiplication laws in $G^{(p)}$ are encoded by an operator product algebra of the symmetry group operators.  For instance in the abstract notation \eqref{symops} above we have 
\begin{equation}\label{defectfuse}
U_{g_{1}}(\Sigma_{d-p-1})U_{g_{2}}(\Sigma_{d-p-1})=U_{g_{1}g_{2}}(\Sigma_{d-p-1})~,
\end{equation}
which generalizes the additive nature of conserved charges.  We can also view this as a rule for what happens when the defects $U_{g_{1}}$ and $U_{g_{2}}$ intersect: they produce $U_{g_{1}g_{2}}$.

\begin{figure}[t]
\centering
\begin{tikzpicture}
\draw [thick, red!90!black] (0,-2) to (0,2); % vertical line
\draw [thick, red!90!black] (0,0) arc (125:155.3:4.5); % wavy line
\draw [thick, red!90!black, dashed, line cap=round] (0,0) arc (-55:-31.7:3.8); 
\draw [thick, red!90!black] (1.5,2.19) arc (-14:-31.5:3.8);
\draw [thick, blue!80!black, decoration = {markings, mark=at position .6 with {\arrow[scale=1.5,rotate=0]{stealth}}}, postaction=decorate] (0,0) arc (-135:-150:6.5); % blue line
\draw (0,2) to (3,-.5) to (3,-4.5) to (0,-2); % plane in front
\draw ([shift=(-130:6)] 0,8) arc (-130:-50:6);
\draw ([shift=(-130:6)] 0,4) arc (-130:-90:6); \draw [dashed, line cap=round] ([shift=(-90:6)] 0,4) arc (-90:-60:6); \draw ([shift=(-60:6)] 0,4) arc (-60:-50:6);
\draw ([shift=(-130:6)] 0,4) to ++(0,4); \draw ([shift=(-50:6)] 0,4) to ++(0,4);
\draw [dashed, dash phase = -2, line cap=round] (-1.5,-1.81) to ++(-1,1.8); \draw (1.5,2.19) to ++(-1,1.8); % plane in back
\draw [dashed, line cap=round] (-2.5,-0.01) to ([shift=(53.1:2.61)] -2.5,-0.01);
\draw [line cap=round] ([shift=(53.1:2.61)] -2.5,-0.01) to ([shift=(53.1:5)] -2.5,-0.01);
\node at (-3.5,2.5) {$\bfg$}; \node at (.5,3.3) {$\bfh$}; \node at (3.4,2.5) {$\bfk$}; \node at (2.4,-3.3) {$\bfg\bfh\bfk$}; % labels
\node at (-.7,-1.5) {\small $\bfg\bfh$}; \node at (0.9,1.75) {\small $\bfh\bfk$};
\node at (-1.3,.2) {\footnotesize $\beta(\bfg, \bfh,\bfk)$};
\filldraw (0,0) circle [radius=.05];
\end{tikzpicture}
\caption{A junction where zero-form symmetry defects of type $\bfg$, $\bfh$, $\bfk$, $\bfg\bfh\bfk\in G$ meet in codimension three. This configuration is generic in spacetime dimension three and above.  The junctions of three codimension-one defects are in red, and their intersection is the black point. At the codimension-three intersection, a one-form symmetry defect $\beta(\bfg,\bfh,\bfk)$ emanates, signaling the 2-group symmetry. In $d$ dimensions, all objects span the remaining $d-3$ dimensions.  
\label{fig: 2-group 3D}}
\end{figure}
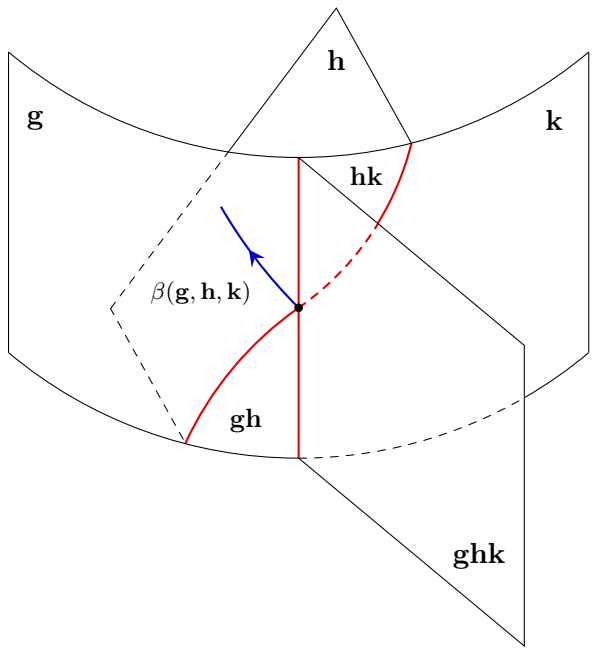

The fusion rule \eqref{defectfuse} encodes only the simplest possible collision of symmetry group operators. More generally we must permit multiple defects to intersect in generic configurations.  When this happens we can encounter defects of various form degrees leading to a higher-group global symmetry.  An example taken from \cite{Benini:2018reh} is shown in Figure \ref{fig: 2-group 3D}.  When the resulting algebra of symmetry defects involves non-trivial intersections with form degrees which are not all equal we say there is higher-group global symmetry. 

Finally, we can also describe higher-group global symmetry in terms of background fields, which is the method that we employ below.  We consider a collection of symmetries and background fields $A^{(1)}, A^{(2)}, \cdots A^{(\ell)}$ that are sources for the higher-form symmetries of different degrees as in \eqref{sourcecurr}.  Each such background field is subject to a gauge transformation as in \eqref{vgauge}, but now we allow for mixing of the gauge transformation between background fields of differing form degree:
\begin{equation}
A^{(p+1)}\longmapsto A^{(p+1)}+d\Lambda^{(p)}+\sum_{k< p} \alpha_{k}(A^{(i)})\Lambda^{(k)}+\text{Schwinger~terms}~.
\end{equation}
Here, the sum is over $k< p$ mirroring the restrictions on $p_{3}$ in equation \eqref{hgrpcur}.  Meanwhile the Schwinger terms above refers to possible pieces of the gauge transformation rules that are non-linear in the gauge parameters.  This sort of mixed gauge transformation rule involving forms of differing degrees is familiar from the Green-Schwarz mechanism \cite{Green:1984sg}.  In that context one encounters such mixed transformations for dynamical fields whereas here it occurs for background fields and signifies a higher-group global symmetry.  Crucially, as we see in examples below, this means that the gauge invariant fluxes for background fields are modified to include Chern-Simons-like corrections constructed out of the other gauge fields.

\subsection{The Symmetries of Axion-Maxwell Theory}
\label{axmaxsymsec}

Let us now describe in more detail the symmetry structure of axion-Maxwell theory, i.e.\ an axion coupled to an abelian gauge field.  We first ignore the coupling between $a$ and the gauge fields and then incorporate modifications due to the interactions. The symmetries and higher-group structure present in these theories has also been discussed in  \cite{Hidaka:2020iaz, Hidaka:2020izy}.

The axion field $a$ has two natural symmetries:  A $U(1)^{(0)}$ shift symmetry and a $U(1)^{(2)}$ two-form symmetry which acts on strings.  The associated currents are respectively:\footnote{The normalization of $J_{1}$ can be checked by noting that the charged operators are $\exp( i n a/f)$ for integer $n$.}
\begin{equation}\label{Jax}
*J_{1}=i f*da~, \hspace{.5in}*J_{3}=\frac{1}{2\pi f}da~.
\end{equation}
The choice of vacuum for $a$ spontaneously breaks the shift symmetry.  Meanwhile the two-form symmetry is always preserved.  The action coupled to background fields $(A^{(1)}, A^{(3)})$ is written as:
\begin{equation}
S=\frac{1}{2}\int(da-f A^{(1)})\wedge *(da- f A^{(1)})+\frac{i}{2\pi f}\int  A^{(3)}\wedge da~.
\end{equation}
In particular, the coupling of the axion to $A^{(1)}$ is standard for a Goldstone mode and under gauge transformations of $A^{(1)}$, $a$ also shifts to ensure the action is invariant.\footnote{In the presence of both $A^{(1)}$ and $A^{(3)}$ the action is not fully invariant under background gauge transformations but instead has a mixed 't Hooft anomaly characterized by an inflow action
\begin{equation}
S_{\mathrm{inflow}}=\frac{i}{2\pi}\int_{X} dA^{(1)}\wedge A^{(3)}~,
\end{equation}
where $X$ is a five-dimensional manifold with boundary the physical spacetime $Y$.  See e.g.\ \cite{Cordova:2019jnf, Cordova:2019bsd} and references therein for an overview of anomaly inflow.
}

Now let us consider an abelian gauge field $C$ with gauge invariant field strength $F=dC$.  We first focus on the case without matter and later generalize to include charged matter fields.  First, in the case of free Maxwell theory without coupling to $a$, there are 1-form symmetries $U(1)_{e}^{(1)}\times U(1)_{m}^{(1)}$, where the subscripts indicate the common association of these as ``electric" and ``magnetic" respectively.  The associated conserved currents are:
\begin{equation}\label{JEM}
*J_{2,e}=\frac{i}{g^{2}}*F~, \hspace{.5in} *J_{2,m}=\frac{F}{2\pi}~.
\end{equation}
It is straightforward to couple these symmetries to background two-form gauge fields $(B_{e}^{(2)}, B_{m}^{(2)})$:
\begin{equation}
S=\frac{1}{2g^{2}}\int (F-B_{e}^{(2)})\wedge * (F-B_{e}^{(2)})+i \int B_{m}^{(2)}\wedge \frac{F}{2\pi}~.
\end{equation}
Under gauge transformations of $B_{e}^{(2)},$ the dynamical gauge field $F$ also shifts indicating that the photon is the Goldstone mode for the electric 1-form symmetry.

We now couple the axion to the Maxwell field leading to the action (without background fields):
\begin{equation}
S=\frac{1}{2}\int da \wedge *da+\frac{1}{2g^{2}}\int F\wedge *F-\frac{i K}{8\pi^{2}f}\int a F\wedge F~,
\end{equation}
Here $K\in \mathbb{Z}$ is a discrete coupling constant.  It is quantized to ensure that the action respects periodicity of the axion.  For some purposes below it is also useful to write the action in way that more manifestly respects both the periodicity of the action and gauge invariance.  This can be done by choosing an auxiliary five-manifold $X$ whose boundary is the physical four-dimensional spacetime $Y$ and writing 
\begin{equation}\label{5dform}
\frac{i K}{8\pi^{2}f}\int_{Y} a F\wedge F=\frac{i K}{8\pi^{2}f}\int_{X} da\wedge F\wedge F~,
\end{equation}
where now the quantized coupling $K$ ensures that the action does not depend on the choice of $X$.  For further mathematical details about the definition of the axion interaction see \cite{Cordova:2019uob}.

The interaction term in the action leads to a modification of the equations of motion.  It is convenient to describe this in terms of the currents introduced in equation \eqref{Jax} and \eqref{JEM}:
\begin{equation}\label{aemeom}
d*J_{3}=0~, \hspace{.2in}d*J_{2,m}=0~, \hspace{.2in}d*J_{1}=\frac{K}{8\pi^{2}}F\wedge F~, \hspace{.2in}d*J_{2,e}= -\frac{K}{4\pi^{2} f}d(aF)~.
\end{equation}
The non-zero right-hand-side of $d*J_{1}$ and $d*J_{2,e}$ above implies that in axion electrodynamics, these symmetries are broken.  In fact, the as we will see these symmetries are broken to discrete groups controlled by $K$:
\begin{equation}
U(1)^{(0)}\longrightarrow \mathbb{Z}_{K}^{(0)}~, \hspace{.2in}U(1)_{e}^{(1)}\longrightarrow \mathbb{Z}_{K}^{(1)}~.
\end{equation}
In particular, in the case of minimal coupling, $K=1$, these symmetries are completely broken.

To demonstrate the above, we need to define topological operators generating discrete transformations.  Those generating the discrete $\mathbb{Z}_{K}^{(0)}$ axion shift symmetry are given by:
\begin{equation}
U_{\ell}(\Sigma_{3})=\exp\left(\frac{2\pi i \ell}{K}\int_{\Sigma_{3}} *J_{1}-\frac{K}{8\pi^{2}}C \wedge F\right)~, \hspace{.2in} \ell \in \mathbb{Z}~.
\end{equation}
By the equations of motion, the argument of the integral is formally closed.  However, since it involves the dynamical gauge field $C$ explicitly it is not manifestly gauge invariant.  The quantization condition $\ell\in \mathbb{Z}$ ensures that the Chern-Simons correction term is well-defined.

Similarly, the symmetry operators for the discrete electric one-form symmetry $\mathbb{Z}_{K}^{(1)}$ are:
\begin{equation}
U_{\ell}(\Sigma_{2})=\exp\left(\frac{2\pi i \ell}{K}\int_{\Sigma_{2}}*J_{2,e}+\frac{K}{4\pi^{2} f}a\,  F\right)~, \hspace{.2in} \ell \in \mathbb{Z}~.
\end{equation}
Again the equations of motion \eqref{aemeom} imply that the integrand is closed so that the operator is topological and the quantization $\ell \in \mathbb{Z}$ implies that the operator is well-defined under the shift redundancy of the axion $a\rightarrow a+2\pi f$.

\subsubsection{Coupling to Background Fields}

We now explain the coupling of axion electrodynamics to background fields.  To do so, we must introduce the appropriate backgrounds for the discrete symmetries $\mathbb{Z}_{K}^{(0)}$ and $\mathbb{Z}_{K}^{(1)}$.  These are discrete gauge fields.  We represent them by standard gauge fields obeying constraints:
\begin{equation}\label{discretegs}
K A^{(1)}=d\lambda^{(0)}~, \hspace{.2in}K B^{(2)}_{e}=d\lambda^{(1)}~,
\end{equation}
where $\lambda^{(i)}$ above are gauge parameters with the periods of $d\lambda^{(i)}$ valued in $2\pi \mathbb{Z}$. Thus the background fields are flat gauge fields with holonomies that are $K$-th roots of unity.  Mathematically one can also view the gauge invariant data in these background gauge fields in terms of cohomology classes.  Under gauge transformation:
\begin{eqnarray}
A^{(1)}\longmapsto A^{(1)}+d\Lambda^{(0)}~, &&\lambda^{(0)}\longmapsto \lambda^{(0)}+K\Lambda^{(0)}~,\\
B_{e}^{(2)}\longmapsto B_{e}^{(2)}+d\Lambda_{e}^{(1)}~, &&\lambda^{(1)}\longmapsto \lambda^{(1)}+K\Lambda_{e}^{(1)}~.\nonumber
\end{eqnarray}
Thus the invariant information is encoded in
\begin{equation}
\left[\frac{K}{2\pi}A^{(1)}\right]\in H^{1}(Y,\mathbb{Z}_{K})~, \hspace{.2in}\left[\frac{K}{2\pi}B_{e}^{(2)}\right]\in H^{2}(Y,\mathbb{Z}_{K})~,
\end{equation}
where the square brackets indicate the gauge equivalence class, and $Y$ is spacetime.

At the linearized level, it is straightforward to couple the symmetries to background fields by introducing couplings of the form \eqref{sourcecurr}.  At the non-linear level this is more subtle due to the periodicity of the axion and the appearance of Chern-Simons terms.  A simple strategy is to recall that the axion $a$ and the gauge field $C$ must still shift under the discrete symmetries.  To account for the possible non-linear mixing between background fields we write the coupling to the $U(1)^{(2)}$ two-form symmetry of the axion and $U(1)^{(1)}$ magnetic symmetry of the gauge field in terms of field strengths:
\begin{equation}
S\supset \frac{i}{2\pi f}\int a\, G^{(4)}-\frac{i}{2\pi}\int H^{(3)}\wedge C~,
\end{equation}
where $G^{(4)}$ and $H^{(3)}$ are gauge invariant field strengths:  
\begin{equation}\label{linear}
G^{(4)}=dA^{(3)}+\cdots~, \hspace{.2in}H^{(3)}=dB^{(2)}_{m}+\cdots~,
\end{equation}
and the omitted terms above indicate possible corrections that are non-linear in the background fields and will be determined below.  Additionally, we find it convenient to use the five-dimensional presentation of \eqref{5dform}.  

With these preliminaries the action coupled to all background fields is simply:
\begin{eqnarray}\label{aemsfull}
S&=&\frac{1}{2}\int_{Y} (da-fA^{(1)})\wedge *(da-f A^{(1)})+\frac{1}{2g^{2}}\int_{Y} (F-B_{e}^{(2)})\wedge * (F-B_{e}^{(2)}) \\
& + & \frac{i}{2\pi f}\int_{Y} a\, G^{(4)}-\frac{i}{2\pi}\int_{Y} H^{(3)}\wedge C-\frac{i K}{8\pi^{2}f}\int_{X} (da-fA^{(1)})\wedge (F-B_{e}^{(2)})\wedge (F-B_{e}^{(2)})~. \nonumber
\end{eqnarray}
Written as above, it is transparent that the theory is invariant under all background gauge transformation and enjoys the correct linear couplings to backgrounds.\footnote{Here we are ignoring shifts by terms involving only background fields i.e.\ 't Hooft anomalies discussed below.}  However, superficially it would seem to depend on the explicit choice of five-manifold $X$.  Demanding that this dependence drop our for all terms that involve dynamical fields fixes the non-linear form of the field strengths in \eqref{linear} to be:\footnote{Mathematically, $\frac{1}{4\pi}B^{(2)}_{e} \wedge B^{(2)}_{e}$ is the Pontryagin square of $B_{e}$ embedded into a continuous variables as in \eqref{discretegs}.}
\begin{equation}\label{G4amaxform}
G^{(4)}=dA^{(3)}+\frac{K}{4\pi}B^{(2)}_{e} \wedge B^{(2)}_{e}~,\hspace{.2in}H^{(3)}=dB_{m}^{(2)}+\frac{K}{2\pi}A^{(1)}\wedge B_{e}^{(2)}~.
\end{equation}
Notice that in general when $B^{(2)}_{e}$ and $A^{(1)}$ are activated, the fluxes $G^{(4)}$ and $H^{(3)}$ are fractional. Moreover, the non-linear terms above mean that the gauge transformations of background fields must also contain non-linear corrections as in the Green-Schwarz mechanism \cite{Green:1984sg}.  In particular this means that while the gauge transformations for $A^{(1)}$ and $B_{e}^{(2)}$ are standard:
\begin{equation}
A^{(1)}\longmapsto A^{(1)}+d\Lambda^{(0)}~, \hspace{.2in}B^{(2)}_{e}\longmapsto B^{(2)}_{e}+d\Lambda^{(1)}_{e}~,
\end{equation}
those of the other background fields contain contain non-linear terms:\footnote{Below we omit possible total derivatives.  As usual in descent these can be absorbed into choices of local counterterms.}
\begin{eqnarray}
A^{(3)}& \longmapsto & A^{(3)}+d\Lambda^{(2)}-\frac{K}{2\pi} B_{e}^{(2)}\wedge d\Lambda_{e}^{(1)}-\frac{K}{4\pi}\Lambda_{e}^{(1)}\wedge d\Lambda_{e}^{(1)}~,\\
B_{m}^{(2)}& \longmapsto & B^{(2)}_{m}+d\Lambda^{(1)}_{m}+\frac{K}{2\pi}A^{(1)}\wedge \Lambda_{e}^{(1)}-\frac{K}{2\pi}B_{e}^{(2)}\, \Lambda^{(0)}+\frac{K}{2\pi}d\Lambda^{(0)}\wedge \Lambda_{e}^{(1)}~.
\end{eqnarray} 
These mixed gauge transformations mean that the symmetries of axion-Maxwell theory form a non-trivial higher-group.  Specifically, in the terminology of section \ref{hgreview}  this is a three-group.  As noted in section \ref{seci} this means that at loci where the one-form symmetry defects intersect we find localized $G^{(4)}$ flux, while at the loci where the defects of $\mathbb{Z}_{K}^{(0)}$ and $\mathbb{Z}_{K}^{(1)}$ intersect we find $H^{(3)}$ flux.  For more details on this algebra of symmetry defects see also \cite{Hidaka:2020izy}.

To conclude this section, let us also note that we can immediately also determine the 't Hooft anomaly from the action \eqref{aemsfull}.  As usual the anomaly is conveniently encoded in an five-dimensional local topological action for the background fields which encodes the anomaly via inflow: 
\begin{equation}\label{inflow1}
S_{\mathrm{inflow}}=\frac{i}{2\pi}\int_{X} G^{(4)}\wedge A^{(1)}+\frac{i}{2\pi}\int_{X}H^{(3)}\wedge B_{e}^{(2)}~.
\end{equation}
This is the appropriate generalization of the separate anomalies of Maxwell theory and the axion theory.

\subsubsection{Extension to Axion-QED}

Let us now discuss the extension of this analysis to theories with matter fields.  We include $N_{F}$ massive fermion flavors of electric charge one.  (Similar analysis applies to theories with charged bosons.) The action is thus:\begin{equation}
S=\frac{1}{2}\int da \wedge *da+\frac{1}{2g^{2}}\int F\wedge *F-\frac{i K}{8\pi^{2}f}\int a F\wedge F+\int i\overline{\Psi}\slashed{D}\Psi +i\overline{\tilde{\Psi}}\slashed{D}\tilde{\Psi} +(m \tilde{\Psi}\Psi+c.c)~,
\end{equation}
where $\Psi$ and $\tilde{\Psi}$ are Weyl fermions of opposite electric charge and the flavor indices on the fermions are suppressed.

The $\mathbb{Z}_{K}^{(0)}$ shift symmetry of the axion is present as before, as is the $U(1)^{(1)}_{m}$ magnetic one-form symmetry and the $U(1)^{(2)}$ two-form axion string symmetry.  However, the presence of the charged matter fields means that Wilson lines can be screened and hence the $\mathbb{Z}_{K}^{(1)}$ one-form symmetry is broken. Instead, in this model we claim that there is a new zero-form flavor global symmetry given by 
\begin{equation}\label{quotient}
\frac{SU(N_{f})^{}}{\mathbb{Z}_{L}}^{(0)}~, \hspace{.5in}L=\mathrm{gcd}(K,N_{f})~.
\end{equation}
The fact that there is a flavor symmetry, which locally (i.e.\ at the Lie algebra level) is of the form $SU(N_{f})$ is obvious from the presence of multiple flavors and the invariant mass.  What remains to be explained is the $\mathbb{Z}_{L}$ quotient above.

The precise meaning of the quotient is that all operators are acted on trivially by the $\mathbb{Z}_{L}$ subgroup of the center of $SU(N_{f})$.  If we look at the sector of local operators that we can build out of $\Psi$'s alone this is clear: every gauge invariant local operator contains equal numbers of $\Psi$'s and $\overline{\Psi}$'s and hence is neutral under the entire $\mathbb{Z}_{N_{f}}$ center of $SU(N_{f})$.  

To see that the quotient is in fact only by $\mathbb{Z}_{L}$ and not $\mathbb{Z}_{N_{f}}$ we examine an axion domain wall across which $a$ jumps by $2\pi f$.  The coupling between $a$ and $F$ then leads to a Chern-Simons theory at level $K$ along the wall.  In particular, this means that monopole operators embedded in the wall carry electric charge $-K$.  We can dress such operators with $K$ modes of $\Psi$ to obtain a gauge invariant field configuration.\footnote{See \cite{Cordova:2017kue} and references therein for a review of the quantum numbers of monopole operators.}  This extended configuration of operators transforms with charge $K$ under the central $\mathbb{Z}_{N_{f}}$ subgroup of $SU(N_{f})$.  This means that while $\mathbb{Z}_{N_{f}}$ in general acts non-trivially on such a dressed monopole, any $\mathbb{Z}_{L}$ subgroup of $\mathbb{Z}_{N_{f}}$ where $L$ divides $K$ acts trivially.  The largest such $L$ is $\mathrm{gcd}(K,N_{f}).$ 

We now proceed to couple the theory to background fields for these global symmetries. The background fields are:
\begin{equation}
\mathbb{Z}_{K}^{(0)}\longleftrightarrow A^{(1)}~, \hspace{.2in}\frac{SU(N_{f})^{}}{\mathbb{Z}_{L}}^{(0)}\longleftrightarrow X^{(1)}~, \hspace{.2in}U(1)^{(1)}_{m}\longleftrightarrow B^{(2)}_{m}~, \hspace{.2in}U(1)^{(2)}\longleftrightarrow A^{(3)}~.
\end{equation}
An important role will be played by the $\mathbb{Z}_{L}$ quotient of the flavor symmetry above.  This quotient means that there are more possible backgrounds available than with the naive $SU(N_{f})$ global symmetry alone.  The additional backgrounds have a topological invariant, the second Steifel-Whitney class $w_{2}\in H^{2}(Y,\mathbb{Z}_{L})$, which one can view as a discrete magnetic flux.  In many ways, this flux plays an analogous role to the electric one-form symmetry background field in the theory without matter.  To emphasize this, we define a dependent background field $B^{(2)}_{e}$ 
\begin{equation}
B^{(2)}_{e}\equiv \frac{2\pi w_{2}}{L}~,
\end{equation}
where $B^{(2)}_{e}$ behaves as a discrete $\mathbb{Z}_{L}$-valued two-form gauge field as in \eqref{discretegs}.  Importantly, when $B_{e}^{(2)}$ is non-zero, the fluxes of the dynamical gauge field become fractional as:
\begin{equation}
\int_{\Sigma_{2}}(F-B_{e}^{(2)})\in 2\pi \mathbb{Z}~.
\end{equation}
This is because the $\mathbb{Z}_{L}$ subgroup defining the quotient in \eqref{quotient} is also a subgroup of dynamical $U(1)$ gauge group.  So in configurations with discrete magnetic flavor flux, the effective gauge group is $U(1)/\mathbb{Z}_{L}$ and the dynamical flux must also be fractional.

With these preliminaries, we can write the action coupled to backgrounds as:
\begin{eqnarray}\label{aqedsfull}
S&=&\frac{1}{2}\int_{Y} (da-fA^{(1)})\wedge *(da-f A^{(1)})+\frac{1}{2g^{2}}\int_{Y} F\wedge * F+\int_{Y} i\overline{\Psi}(\slashed{D}+\slashed{X}^{(1)})\Psi+i\overline{\tilde{\Psi}}(\slashed{D}-\slashed{X}^{(1)})\tilde{\Psi}  \nonumber \\
& + & (m \tilde{\Psi}\Psi+c.c)+\frac{i}{2\pi f}\int_{Y} G^{(4)}\wedge a-\frac{i}{2\pi}\int_{Y} H^{(3)}\wedge C-\frac{i K}{8\pi^{2}f}\int_{X} (da-fA^{(1)})\wedge F\wedge F~. 
\end{eqnarray}
Demanding as before that the action depends only on the dynamical variables in the physical four-dimensional spacetime results in the gauge invariant field strengths:
\begin{equation}\label{h3formaqed}
G^{(4)}=dA^{(3)}+\frac{K}{4\pi}B_{e}^{(2)}\wedge B_{e}^{(2)}~, \hspace{.5in}H^{(3)}=dB_{m}^{(2)}+\frac{K}{2\pi}A^{(1)}\wedge B^{(2)}_{e}~.
\end{equation}
Thus, as before we find a higher-group structure, now involving the flavor symmetry, whenever $L=\gcd(K,N_{f})$ is larger than one.  We also find an 't Hooft anomaly in this theory with inflow action again given by \eqref{inflow1}.

\subsection{The Symmetries of Axion-Yang-Mills}
\label{aymsymsec}

We now discuss axion-Yang-Mills.  For simplicity, we restrict ourselves to gauge group $SU(N)$.  The higher-group symmetries we uncover follow directly from the results of \cite{Seiberg:2018ntt, Cordova:2019uob}.  To begin let us describe symmetries of the Yang-Mills sector independent of the axion.  This theory has a one-form symmetry $\mathbb{Z}_{N}^{(1)}$ related to the center of the gauge group \cite{Gaiotto:2014kfa}.  Physically this symmetry is present because the only dynamical fields are the adjoint gluons which are neutral under the center of the gauge group. Thus, one cannot screen Wilson lines in representations which are charged under the center of the gauge group. This leads to a quantum number, conserved modulo $N,$ that is carried by Wilson lines.  In other words a one-form symmetry.  In many ways this $\mathbb{Z}_{N}^{(1)}$ is a discrete analog of the $U(1)^{(1)}_{e}$ symmetry of Maxwell theory.  By contrast, the $U(1)^{(1)}_{m}$ symmetry has no discrete avatar in $SU(N)$ Yang-Mills.\footnote{This situation is changed for other global forms of the gauge group.  For instance in $SU(N)/\mathbb{Z}_{N}$ Yang-Mills, there is no one-form charge carried by the Wilson lines, but instead the 't Hooft lines are charged under a $\mathbb{Z}_{N}^{(1)}$. }

To couple the $SU(N)$ theory to an appropriate background field we follow the procedure described in \cite{Gaiotto:2017yup}.  The background is again represented by a constrained gauge field $B^{(2)}$ obeying
\begin{equation}
NB^{(2)}=d\lambda^{(1)}~,\hspace{.2in} \left[\frac{N}{2\pi}B^{(2)}\right]\in H^{2}(Y,\mathbb{Z}_{N})~.
\end{equation}
We first promote the dynamical gauge field to be a $U(N)$ connection (for simplicity we denote this by the same symbol $C$ now valued in the adjoint of $U(N)$).  We then introduce a Lagrange multiplier field $\varphi$ which constrains the trace of $C$ to be given by the background source:
\begin{equation}\label{Lagrange}
S\supset \frac{1}{2\pi} \int \varphi (\mathrm{Tr}(F)-NB^{(2)})~.
\end{equation}
Thus on the solutions to the equation of motion, the abelian part of the field strength is fixed by $B^{(2)}$ and so $F=B^{(2)} \mathbf{1}+\tilde{F}$, with $\mathbf{1}$ the $N\times N$ identity matrix and $\tilde{F}$ traceless.  As in the Maxwell example, $C$ shifts under the one-form symmetry.  This means that under gauge transformations:
\begin{equation}
B^{(2)}\longmapsto B^{(2)}+d\Lambda^{(1)}~, \hspace{.2in}\lambda^{(1)}\longmapsto \lambda^{(1)}+N\Lambda^{(1)}~, \hspace{.2in}C\longmapsto C+\Lambda^{(1)}\mathbf{1}~,
\end{equation}
so the interaction \eqref{Lagrange} is invariant under background gauge transformations.  Notice in particular that if $B^{(2)}$ vanishes, $F$ is traceless and the remaining background gauge redundancy can be used to shift $C$ by an arbitrary flat $U(1)$ gauge field.  Thus, we are reduced to the original $SU(N)$ theory.  More generally, a $U(N)$ Yang-Mills theory with the interaction \eqref{Lagrange} has the same number of dynamical degrees of freedom as $SU(N)$ Yang-Mills.

With these remarks it is straightforward to write the full action coupled to $B^{(2)}$ as:
\begin{equation}
S= \int \frac{\varphi}{2\pi} (\mathrm{Tr}(F)-NB^{(2)})+\frac{1}{g^{2}} \mathrm{Tr}\left[(F-B^{(2)}\mathbf{1})\wedge *(F-B^{(2)}\mathbf{1})\right]
+\frac{i\theta}{8\pi^{2}}\mathrm{Tr}\left[(F-B^{(2)}\mathbf{1})\wedge (F-B^{(2)}\mathbf{1})\right]~.
\end{equation}
As it will be important in the following, let us examine the $\theta$ term in more detail.  We recall the formula for the integral second Chern number of a $U(N)$ gauge field:
\begin{equation}
c_{2}=\frac{1}{8\pi^{2}}\int \mathrm{Tr}(F)\wedge \mathrm{Tr}(F)-\mathrm{Tr}(F\wedge F)~, \hspace{.3in}c_{2}\in \mathbb{Z}
\end{equation}
Then, using the constraint arising from \eqref{Lagrange} we have:
\begin{eqnarray}\label{thetaterms}
\frac{i\theta}{8\pi^{2}}\mathrm{Tr}\left[(F-B^{(2)}\mathbf{1})\wedge (F-B^{(2)}\mathbf{1})\right]& = & \frac{i\theta}{8\pi^{2}}\int \left[\left(\mathrm{Tr}(F\wedge F)-\mathrm{Tr}(F)\wedge \mathrm{Tr}(F)\right) +N(N-1)B^{(2)}\wedge B^{(2)}\right] \nonumber \\
& = & -i\theta c_{2}+\frac{i\theta N(N-1)}{8\pi^{2}}\int B^{(2)}\wedge B^{(2)}~.
\end{eqnarray}
In particular, since $B^{(2)}$ has periods that are $N$-th roots of unity, we see that in the presence of background $B^{(2)},$ the periodicity of $\theta$ is enlarged to $\theta \sim \theta+2\pi N$.  This effect will be responsible for the intricate symmetry structure present in axion Yang-Mills.\footnote{Here, we are restricting ourselves to spacetimes which are spin, appropriate to later applications involving fermions.  More generally, on a non-spin manifold the periodicity of $\theta$ is $2\pi N$ for $N$ odd and $4\pi N$ for $N$ even.}

It is clarifying to describe the manipulations above using characteristic classes.  We are making use of the fact that $U(N)\cong \frac{U(1)\times SU(N)}{\mathbb{Z}_{N}}$.  The $\mathbb{Z}_{N}$ quotient means that the allowed gauge bundles of $U(N)$ are more general than products of $U(1)$ bundles and $SU(N)$ bundles.  Instead, a $U(N)$ bundle can be described as a $U(1)/\mathbb{Z}_{N}$ bundle and a $SU(N)/\mathbb{Z}_{N}$ bundle subject to a constraint.  This means that the $U(1)$ flux may be fractional (an $N$-th root of unity) while the discrete magnetic flux of $SU(N)/\mathbb{Z}_{N}$, mathematically, the second Steifel-Whitney class $w_{2}\in H^{2}(Y,\mathbb{Z}_{N}),$ may be non-zero as long as these quantities are related as:
\begin{equation}
\frac{1}{2\pi}\mathrm{Tr}(F)=w_{2}=\left[\frac{N}{2\pi}B^{(2)}\right]\in H^{2}(Y,\mathbb{Z}_{N})~.
\end{equation}
In particular, as is well-known, bundles of $SU(N)/\mathbb{Z}_{N}$ have fractional instanton number which is precisely encoded by formula \eqref{thetaterms}.\footnote{As in \eqref{G4amaxform}, we can express this in terms of the Pontryagin square, with the fractional part of the instanton number being given by $\frac{N-1}{2N}\mathcal{P}(w_{2})$.}

\subsubsection{Background Fields for Axion Yang-Mills}

We now describe the symmetries of axion-Yang-Mills.  The analysis is similar to that of axion-Maxwell theory.  In the absence of background fields, the action is
\begin{equation}\label{aymact}
\frac{1}{2}\int da \wedge *da +\frac{1}{g^{2}}\int \mathrm{Tr}(F\wedge *F)-\frac{iK}{8\pi^{2}f}\int a \mathrm{Tr}(F\wedge F)~,
\end{equation}
where above, the coupling constant $K$ is a positive integer.  There are three symmetry groups of interest: $\mathbb{Z}_{K}^{(0)}$, $\mathbb{Z}_{N}^{(1)}$, and $U(1)^{(2)}$.

We now activate background fields and account for possible non-linear corrections to the gauge transformations and fluxes.  As before, this is most clear by writing the axion term as a coupling on a five-manifold $X$ with boundary the physical spacetime $Y$:
\begin{eqnarray}
S & = & \frac{1}{2}\int_{Y}(da-fA^{(1)})\wedge*(da-fA^{(1)})+\int_{Y} \frac{\varphi}{2\pi} (\mathrm{Tr}(F)-NB^{(2)})+ \frac{i}{2\pi f}\int_{Y}a\, G^{(4)}\\
&+&\frac{1}{g^{2}} \int_{Y}\mathrm{Tr}\left[(F-B^{(2)}\mathbf{1})\wedge *(F-B^{(2)}\mathbf{1})\right] -\frac{iK}{8\pi^{2}f}\int_{X}(da-fA^{(1)})\mathrm{Tr}\left[(F-B^{(2)}\mathbf{1})\wedge (F-B^{(2)}\mathbf{1})\right]~. \nonumber
\end{eqnarray}
Demanding that, modulo possible anomalies, the action depends only on the physical spacetime $Y$ implies that the gauge invariant field strength must be:
\begin{equation}\label{aymg4}
G^{(4)}=dA^{(3)}+\frac{KN(N-1)}{4\pi} B^{(2)}\wedge B^{(2)}~.
\end{equation}
In particular, this means that while the gauge transformations of $A^{(1)}$ and $B^{(2)}$ are standard, the gauge transformation rules for $A^{(3)}$ are more interesting:
\begin{equation}
A^{(3)}\longmapsto A^{(3)}+d\Lambda^{(2)}-\frac{KN(N-1)}{2\pi}\Lambda^{(1)}\wedge B^{(2)}-\frac{KN(N-1)}{4\pi}\Lambda^{(1)}\wedge d\Lambda^{(1)}~.
\end{equation} 
Again, this is an example of a higher-group global symmetry.  

We can also readily read off the anomaly inflow action
\begin{equation}
S_{\mathrm{inflow}}=\frac{i}{2\pi}\int_{X} G^{(4)}\wedge A^{(1)}~.
\end{equation}

\subsubsection{Extension to Axion-QCD}

As a final model of axions, we consider axions coupled to Yang-Mills fields and $N_{f}$ fundamental quarks, all with equal mass. The results follow straightforwardly from the previous examples and our treatment is brief.  The action is given by:
\begin{equation}
\frac{1}{2}\int da \wedge *da +\frac{1}{g^{2}}\int \mathrm{Tr}(F\wedge *F)-\frac{iK}{8\pi^{2}f}\int a \mathrm{Tr}(F\wedge F)+\int i\overline{\Psi}\slashed{D}\Psi+i\overline{\tilde{\Psi}}\slashed{D}\tilde{\Psi} +(m \tilde{\Psi}\Psi+c.c)~,
\end{equation}
where the gauge and flavor indices on the fermions $\Psi$ are suppressed. The presence of fundamental matter means that the one-form symmetry of axion-Yang-Mills is broken.  However, there is a new zero-form flavor symmetry $\frac{U(N_{f})}{\mathbb{Z}_{N}}^{(0)}$ arising from rotating the quarks, where the $\mathbb{Z}_{N}$ quotient is because it is identified with the center of the gauge group.  In summary the symmetry is:
\begin{equation}
\mathbb{Z}_{K}^{(0)}\times \frac{U(N_{f})}{\mathbb{Z}_{N}}^{(0)}\times U(1)^{(2)}~.
\end{equation}

We will now show that these symmetries are extended into a non-trivial higher group whenever $N$ does not divide $K$.  The key feature is that when global symmetry backgrounds that make use of the $\mathbb{Z}_{N}$ quotient are activated, the flux of the dynamical Yang-Mills field is also fractional.

Following \cite{Cordova:2019uob}, we can understand this by looking at the gauge and global group action on the fundamental fermions.  The part that acts faithfully is given by
\begin{equation}
\frac{SU(N)\times U(N_{f})}{\mathbb{Z}_{N}}\cong \frac{SU(N)}{\mathbb{Z}_{N}}\times \frac{SU(N_{f})}{\mathbb{Z}_{N_{f}}}\times \frac{U(1)}{\mathbb{Z}_{M}}~, \hspace{.2in}M=\mathrm{lcm}(N,N_{f})~,
\end{equation}
and above our convention is that the $U(1)$ factor (under which baryons are charged) acts on the fundamental fermions with charge one.  This means that in general, the discrete magnetic flux of the dynamical gauge group $w_{2}(SU(N))\in H^{2}(M,\mathbb{Z}_{N})$ is non-zero, as is that of the flavor group $w_{2}(SU(N_{f}))\in H^{2}(M,\mathbb{Z}_{N_{f}})$ and the fractional part of the $U(1)$ flavor flux (denoted by $Z$ below).  However they are related via: 
\begin{equation}\label{flavres}
\int_{\Sigma_{2}} \frac{Z}{2\pi}=\frac{1}{N}\int_{\Sigma_{2}}w_{2}(SU(N))+\frac{1}{N_{f}}\int_{\Sigma_{2}}w_{2}(SU(N_{f}))+s~, \hspace{.2in}s\in \mathbb{Z}~.
\end{equation}
Similarly, we can also write a formula relating the fractional flux of the $U(1)$ to the discrete magnetic flux $w_{2}(SU(N_{f}))$ for the flavor bundles of $U(N_{f})/\mathbb{Z}_{N}$ as:
\begin{equation}\label{flavconst}
N\int_{\Sigma_{2}} \frac{Z}{2\pi}=\frac{N}{N_{f}}\int_{\Sigma_{2}}w_{2}(SU(N_{f}))+\ell~, \hspace{.2in}\ell\in \mathbb{Z}~.
\end{equation}
By varying the fluxes of the flavor gauge fields subject to \eqref{flavconst}, we can achieve any value of the $w_{2}(SU(N))$ determined via \eqref{flavres}, and therefore a general fractional instanton number as in \eqref{thetaterms}.

To match our previous discussion and notation we introduce a background field $B^{(2)}$ dependent on the zero-form symmetry as:
\begin{equation}
B^{(2)}=\frac{2\pi}{N_{f}}w_{2}(SU(N_{f}))~.
\end{equation}
Then the instanton number is fractional with
\begin{equation}
\frac{1}{8\pi^{2}}\int \mathrm{Tr}(F\wedge F)=\frac{N(N-1)}{8\pi^{2}}(Z-B)\wedge (Z-B)~,
\end{equation}
Hence proceding as before leads to a formula for the gauge invariant four-form field strength coupling to the axion:
\begin{equation}\label{g4aqcdform}
G^{(4)}=dA^{(3)}+\frac{KN(N-1)}{4\pi}(Z-B)\wedge (Z-B)~.
\end{equation}
In particular, the fluxes of $G^{(4)}/2\pi$ are fractional as long as $N$ does not divide $K$.

\section{Higher-Groups and Constraints on Emergence}
\label{secemerge}

The symmetries that we have discussed in the previous section are examples of the general phenomena of higher-group global symmetry described in section \ref{hgreview}.  Here we use this to deduce model independent inequalities on the scales of emergent symmetries in models of axions following the general logic of \cite{Cordova:2018cvg}.

\subsection{Inequalities for Axion-Yang-Mills}
\label{symscalesec}

Consider as an illustrative example the case of axion-$SU(N)$-Yang-Mills described in section \ref{aymsymsec}.  The action is given by \eqref{aymact} and the symmetries are $\mathbb{Z}_{K}^{(0)}\times \mathbb{Z}_{N}^{(1)}\times U(1)^{(2)}$.  The key result is equation \eqref{aymg4} for the gauge invariant four-form field strength associated to the symmetry $U(1)^{(2)}$:
\begin{equation}
G^{(4)}=dA^{(3)}+\frac{KN(N-1)}{4\pi} B^{(2)}\wedge B^{(2)}~,
\end{equation}
where $B^{(2)}$ is the background field for the $ \mathbb{Z}_{N}^{(1)}$ one-form global symmetry.  In particular notice that when $B^{(2)}$ is activated, there is in general also a non-trivial four-form flux $G^{(4)}$ (typically fractional).

Now imagine a renormalization group flow where both the symmetries in question are emergent.  Let $E_{\mathrm{screen}}$ be the scale below which the $\mathbb{Z}_{N}^{(1)}$ is a good symmetry of the effective field theory, and $E_{\mathrm{string}}$ the scale below which $U(1)^{(2)}$ is a good symmetry.  Then below $E_{\mathrm{screen}}$ we can couple our theory to general backgrounds $B^{(2)}$ and hence, since $G^{(4)}$ is also in general activated, we must also be below $E_{\mathrm{string}}$ so that we can activate this background as well.  Therefore we deduce that for any renormalization group trajectory which at low energies flows to axion-Yang-Mills, we have model independent inequalities:
\begin{equation}\label{ineq12}
E_{\mathrm{screen}}\lsim E_{string}~.
\end{equation}
Of course, as remarked in the introduction, the scales $E_{\mathrm{screen}}, E_{\mathrm{string}}$ are only approximately defined along the flow.  Therefore, we interpret the above as the statement that $E_{\mathrm{screen}}$ cannot be parametrically larger than $E_{\mathrm{string}}$.

\subsubsection{Pecci-Quinn Example}

It is useful to make the inequality more precise in the case of an explicit  KSVZ axion model \cite{Kim:1979if,Shifman:1979if}, where the the field $a$ emerges from a Pecci-Quinn mechanism.  In the ultraviolet, we have $SU(N)$ gauge theory with $K$ Weyl fermions in the fundamental ($\psi_+$) and anti-fundamental $(\psi_-$) representation (flavor indices suppressed) and an uncharged complex scalar field $\varphi$. The action is:   
\eq{\label{spq}
S=\int d^4x\Bigg\{ \frac{1}{g^{2}}\Tr[F\wedge *F]+i \bar\psi_\pm  \slashed{D}\psi_\pm +\frac{1}{2}d\varphi\wedge *d\bar{\varphi}-V(\varphi)+\lambda \bar\varphi \psi_+\psi_-+\lambda \varphi \bar\psi_- \bar\psi_+\Bigg\}~.
}
We assume the coupling constants above are small, so that semiclassical analysis is reliable.

At the classical level, there is a $U(1)_{PQ}$ symmetry where the different fields have charges 
\begin{center}
\begin{tabular}{c|c}
Field&$U(1)_{PQ}$ charge\\
\hline
$\psi_\pm$&$ +1$\\
$\varphi$&$+2$
\end{tabular}
\end{center}
At the quantum level, this $U(1)_{PQ}$ is broken by an ABJ anomaly.  Let the scalar potential be:
\eq{
V(\varphi)=m^2 (|\varphi|^2-f^2)^2~, 
}
so that $\varphi$ condenses and gives a mass to the fermions through the Yukawa couplings.  When we flow to energies below the mass of the radial mode of $\varphi$ we will see that this theory is described by the gauge fields and a pseudo-goldstone mode $a$ which is an axion. 

More explicitly, we write the complex scalar field as 
\begin{equation}
\varphi=\rho e^{ia/f}~.
\end{equation}
The radial mode $\rho$ has a mass $m_{\rho}\sim mf$.  Below this scale the action reduces to:
\eq{
S=\int d^4x \Bigg\{ \frac{1}{g^{2}} \Tr[F\wedge *F]+i \bar\psi_\pm \slashed{D}\psi_\pm + \frac{1}{2}da\wedge *da+\lambda fe^{-ia/f} \psi_+\psi_-+\lambda f e^{i a /f}\bar\psi_+\bar\psi_-\Bigg\}~. 
}
Via a chiral rotation of the fermions we can eliminate the $a$-dependence of the fermion mass term.  The fermions are then massive with $m_{\psi}=\lambda f$ and provided this is above the QCD scale so that the model remains weakly coupled we can integrate them out.  However, since the chiral symmetry is anomalous, this chiral rotation generates an axion coupling between $a$ and the remaining gauge fields.  Thus the final action is that of axion-Yang-Mills.  
\begin{equation}
\frac{1}{2}\int da \wedge *da +\frac{1}{g^{2}}\int \mathrm{Tr}(F\wedge *F)-\frac{iK}{8\pi^{2}f}\int a \mathrm{Tr}(F\wedge F)~,
\end{equation}

\begin{figure}
\begin{center}
\label{fig:mexicanHat}
\includegraphics[scale=0.7,clip,trim=4cm 22cm 5cm 1.9cm]{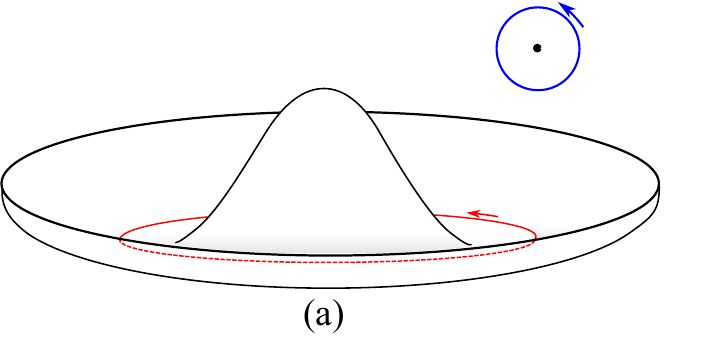}\includegraphics[scale=0.7,clip,trim=4cm 22cm 5cm 1.9cm]{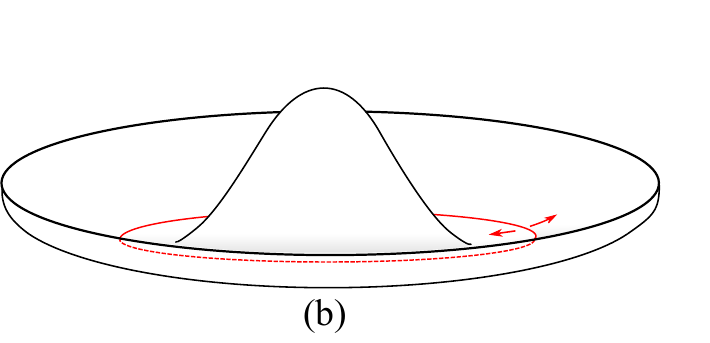}\\
\includegraphics[scale=0.4,clip,trim=4cm 16cm 5cm 8cm]{MexicanHatAxionFig.pdf}
\includegraphics[scale=0.4,clip,trim=4cm 8.5cm 5cm 15.5cm]{MexicanHatAxionFig.pdf}
\includegraphics[scale=0.4,clip,trim=4cm 1cm 5cm 23cm]{MexicanHatAxionFig.pdf}
\end{center}
\caption{This figure illustrates how the axion string can unwind. In (a), we show the winding solution for the axion string along a circle of fixed radius. As we go around the blue circle (a circle linking the axion string (black) in spacetime) the scalar field $\varphi$ winds around the bottom of the mexican hat potential in red. (b) shows the excitations of the radial mode which is activated at the scale $m_\rho \sim m f$. (c-e) shows a process by which we can unwind the scalar field. Here we deform solution in the radial direction over the top of the potential which costs energy $\sqrt{m}f>>m f$. The resulting configuration has no winding -- indicating the decay of the string.  \label{fig:hat}}
\end{figure}

The UV theory has neither a one-form nor a two-form global symmetry. Both are emergent at low energies and are associated to characteristic energy scales.  Specifically:
\begin{itemize}
\item The one-form symmetry is associated to a scale
\begin{equation}
E_{\mathrm{screen}}\approx \lambda f~,
\end{equation}
which is the mass of the fundamental fermions.  Indeed above this scale, the fermions fluctuate and can screen the charged Wilson lines breaking the one-form symmetry $\mathbb{Z}_{N}^{(1)}.$
\item Defining the two-form symmetry scale $E_{\mathrm{string}}$  is more subtle.  There are several scales naturally associated to the emergence of the axion.  First, there is the string tension $T_{\mathrm{string}}^{1/2}=f$.\footnote{\label{StringTen}  
The tension of the axion string can be estimated by approximating a charge $P$ axion string solution as
\eq{
\varphi=f\theta(|x|-\ell_{core}) e^{i P\phi}~,
}
where $\ell_{core}$ is the characteristic size of the string. The tension can be roughly determined by the potential energy:
\eq{
T_{\mathrm{string}}\sim \int d^2x \,m^2(|\varphi|^2-f^2)^2=\pi m^2 f^4 \ell_{core}^2~.
}
Then using $\ell_{core}\sim \frac{1}{m_{\rho}}\sim \frac{1}{mf}$, we find that the tension goes like $T_{\mathrm{string}}\sim f^2$. } There is also the scale $m_{\rho}\approx mf$ where the radial mode begins to fluctuate.  At weak coupling, $m<<1$ these scales are well separated $T_{\mathrm{string}}^{1/2}\sim f>>m_{\rho}\sim mf$.

Note that just above the energy $m_{\rho}$, the scalar potential confines $\varphi$ to valley surrounding the minimum and the scalar field effectively becomes $S^1\times I$-valued where $I$ is an interval.  It is this effective non-trivial topology in the space of fields, i.e.\ the presence of a circle factor, that ensures there is a two form-symmetry and a topologically conserved winding number. In particular just above the scale $m_{\rho}$ the two-form symmetry persists.  As we go to yet higher scales, we eventually reach the energy $E_{\mathrm{string}}$ where the $\rho$ particle can traverse all the way to the origin in field space and break the two-form symmetry.  Physically this means that at this scale there are processes that allow strings to unwind (See Figure \ref{fig:hat}).  This scale is determined by the value of the potential at the origin $V(0)\sim m^2 f^4$ so that
\begin{equation}
E_{\mathrm{string}}\approx \sqrt{m}f~.
\end{equation} 
At weak-coupling $m<<1$ there is thus a hierarchy: 
\eq{
T_{\mathrm{string}}^{1/2}=f~~>>~~E_{\mathrm{string}}=\sqrt{m} f~~>>~~ m_{\rho}=mf~.
}
\end{itemize}   

Having identified the emergent symmetry scales in the problem, we can now apply the general inequality \eqref{ineq12} to deduce
\eq{\label{aYMequality}
\sqrt{m}\gtrsim \lambda \Longleftrightarrow m_{\rho}f \gtrsim m_{\psi}^{2}~,
}
where the left-hand-side expresses the constraint in terms of dimensionless coupling constants, and the right-hand side expresses the same constraint in terms of the physical masses and tensions in the problem.

It is interesting to explore what happens when the inequality \eqref{aYMequality} is violated.  Since the scales in question are all approximate, we consider a parametric limit within a weakly-coupled setting where we retain technical control: $1>>\lambda^{2} >>m$.  We will argue that the Pecci-Quinn field theory \eqref{spq} breaks down.

To begin let us note that the action \eqref{spq} does not define a UV complete theory: at very high energies the scalar quartic coupling runs to a Landau pole.  We interpret the UV action as an effective theory defined with a cutoff scale $\Lambda_{UV}>>f.$  We interpret the inequality \eqref{aYMequality} as a statement about the couplings defined at the scale $\Lambda_{UV}$.  Thus, the inequality is parametrically violated when $\lambda^2(\Lambda_{UV})>>m(\Lambda_{UV})$. In this limit, we will find that the fermions significantly modify the effective scalar potential. 

It is straightforward to compute the effective potential by integrating out the fermions in a background of constant $\varphi$.  (For convenience, one can combine the pair of Weyl fermions $\psi_\pm$ into a single Dirac fermion, and express $\varphi$ in terms of its real and imaginary parts).  Their contribution to the effective scalar potential is then: 
\begin{eqnarray}
\Delta V&=&
\int d\lambda \partial_\lambda \Tr\, \log(-i \slashed{D}+\lambda (\varphi_R+i \gamma^5 \varphi_I))\\
& = &\frac{\lambda^4 N K}{8\pi^2}\left(|\varphi|^4\log\left(\frac{\lambda|\varphi|}{\Lambda_{UV}}\right)-\frac{|\varphi|^4}{4}\right)~. \nonumber
\end{eqnarray}
The full effective potential for the theory is therefore:
\eq{
V_{eff}=\frac{\lambda^4 N K}{8\pi^2}\left(|\varphi|^4\log\left(\frac{\lambda|\varphi|}{\Lambda_{UV}}\right)-\frac{|\varphi|^4}{4}\right)+m^2(|\varphi|^2-f^2)^2~,
}
where all couplings are evaluated at the scale $\Lambda_{UV}$.  

When $\lambda^{2}$ is small compared to $m,$ it is clear that the tree level potential dominates and one finds a minimum near $\langle |\varphi|\rangle=f$ and the previous analysis leading to axion-Yang-Mills is valid.  However, when the quartic coupling is non-negligible the minimum moves drastically. Working in the limit where the cutoff scale is large, $\Lambda_{UV}>>\lambda f$, we find that the implied vacuum is at:
\begin{equation}
\langle |\varphi|\rangle \approx  \frac{\Lambda_{UV}}{\lambda}\exp\left(-\frac{8\pi ^{2}m^{2}}{NK\lambda^{4}}\right)~.
\end{equation} 
In particular when $\lambda^{2}>>m$ so that the inequality \eqref{aYMequality} is violated, the minimum is larger than the cutoff scale $\Lambda_{UV}$.\footnote{Notice that in the limit of interest $\lambda^{2}>>m$ the logarithm is small and we expect the one-loop effective potential to be reliable.}  When this happens the effective field theory breaks down and we lose control.  Indeed, in formulating the effective field theory, we have neglected irrelevant operators in the action suppressed by powers of the cutoff.  For instance, terms in the potential of the form $|\varphi|^{L+4}/\Lambda_{UV}^{L}$ for large $L$ have been discarded.  This analysis is correct provided that the minimum of $|\varphi|$ is small, but when $|\varphi|$ is of order the cutoff the EFT analysis breaks down.  Thus we see that the inequality \eqref{aYMequality}, derived from general considerations of symmetry, is built in to the consistency of the effective field theory.

\subsection{Inequalities for Axion-QCD}

We can apply similar logic to $SU(N)$ axion-QCD which has global symmetry $\mathbb{Z}_{K}^{(0)}\times \left(U(N_{f})/\mathbb{Z}_{N}\right)^{(0)}\times U(1)^{(2)}$.  As found in equation \eqref{g4aqcdform}, the fluxes of $G^{(4)}$ are necessarily activated (and fractional) provided that we activate general backgrounds of $\left(U(N_{f})/\mathbb{Z}_{N}\right)^{(0)}$ that make use of the $\mathbb{Z}_{N}$ quotient. Therefore, we again obtain constraints on the scales of symmetry emergence for any flow that realizes axion-QCD in the IR:
\begin{equation}\label{ineqaqcd}
E_{\mathbb{Z}_{N}} \lsim E_{\mathrm{string}}~.
\end{equation}
Here, the scale $E_{\mathrm{string}}$ is again the string stability scale, while $E_{\mathbb{Z}_{N}}$ is the energy at which operators charged under the $\mathbb{Z}_{N}\subset U(N_{f})$ appear.

As in our analysis of axion-Yang-Mills inequalities we expect that \eqref{ineqaqcd} is enforced by the consistency of any UV effective field theory which flows to axion-QCD at long distances.  As a simple example consider the action:
\begin{eqnarray}
S&=&\int d^4x\Bigg\{ \frac{1}{g^{2}}\Tr[F\wedge *F]+i \bar\psi_\pm  \slashed{D}\psi_\pm +\frac{1}{2}d\varphi\wedge *d\bar{\varphi}-m^2(|\varphi|^2-f^2)^2+\lambda \bar\varphi \psi_+\psi_-+\lambda \varphi \bar\psi_- \bar\psi_+ \nonumber \\
& + &i \bar\Psi_I \slashed{D}\Psi^I+i \overline{\tilde{\Psi}}^I \slashed{D}\tilde\Psi_I+(M \tilde\Psi_I \Psi^I+c.c.)\Big\}~,
\end{eqnarray}
where the fermions $\Psi^I$ ($\tilde\Psi_I$) transform in the fundamental (anti-fundamental) representation of the gauge group $SU(N)$. 

Note that in this EFT the relevant the flavor symmetry is in fact  $U(N_f)^{(0)}$ not $\left(U(N_f)/\IZ_N\right)^{(0)}$. The reason is that we can construct the local, gauge invariant operators 
\be
\CO^I=\psi_1\Psi^I~, \hspace{.5in} \tilde\CO_I=\psi_2 \tilde\Psi_I~,
\ee
which transform non-trivially under $\mathbb{Z}_{N}$.   When we flow to the IR, the Higgs field gives the PQ-fermions a mass.  This reduces the flavor symmetry $U(N_f)\longrightarrow U(N_f)/\IZ_N$.
Therefore, we identify the scale $E_{\mathbb{Z}_{N}}\approx \lambda f$.  Meanwhile, following the discussion of axion-Yang-Mills, the string stability scale is $E_{\mathrm{string}}\approx \sqrt{m}f$.  The inequality \eqref{ineqaqcd} reduces to $\lambda \lsim \sqrt{m}$ and is derived identically to the analysis of axion Yang-Mills above.  Again, when the inequality is parametrically violated the PQ fermions make large contributions to the scalar potential and the EFT breaks down.

\subsection{Inequalities for Axion Electrodynamics}

We can also apply our analysis to axion-electrodynamics.  As discussed in section \ref{axmaxsymsec}, there are several higher group structures to consider when the axion coupling constant $K>1$.  The implications of the higher group extension of $U(1)^{(2)}$ are identical to those of the previous section: the emergence scale $E_{\mathrm{string}}$ must be the largest scale of those involved in the higher group.  

Somewhat more novel is the higher group extension of the $U(1)^{(1)}_{m}$ magnetic one-form symmetry which measures the charges of 't Hooft lines. The gauge invariant field strength derived in \eqref{G4amaxform} and \eqref{h3formaqed} is:
\begin{equation}
H^{(3)}=dB_{m}^{(2)}+\frac{K}{2\pi}A^{(1)}\wedge B^{(2)}_{e}~.
\end{equation}
Here we recall that $A^{(1)}$ is the background field for the $\mathbb{Z}_{K}^{(0)}$ shift symmetry of the axion.  Meanwhile, in axion-Maxwell theory (with no charged matter) $B^{(2)}_{e}$ is the background for the electric $\mathbb{Z}_{K}^{(1)}$ one-form symmetry, while it is a composite due to the quotient in the flavor symmetry $\left(SU(N_{f})/\mathbb{Z}_{L}\right)^{(0)}$ ($L=\gcd(K,N_{f})$) in the axion-QED.  In particular, we see that if both $A^{(1)}$ and $B^{(2)}_{e}$ are present then necessarily $H^{(3)}$ fluxes are also present.  Following similar logic to our analysis of axion-YM in section \ref{symscalesec}, this leads to a constraint on the scales of symmetry emergence.  

In the case of axion Maxwell theory, we let $E_{\mathrm{magnetic}}$ be the scale of emergence of the $U(1)^{(1)}_{m}$ magnetic one-form symmetry.  We also denote by $E_{\mathrm{shift}}$ the scale where the $\mathbb{Z}_{K}^{(0)}$ shift symmetry emerges, and $E_{\mathrm{screen}}$ the scale where the $\mathbb{Z}_{K}^{(1)}$ electric one-form symmetry emerges.  Then we obtain a universal inequality:
 \begin{equation}\label{ineqamax}
\min\{E_{\mathrm{shift}}, ~E_{\mathrm{screen}}\} \lsim E_{\mathrm{magnetic}}~.
\end{equation}
Similarly in axion-QED the role of $E_{\mathrm{electric}}$ is now played by $E_{\mathbb{Z}_{L}}$, above which there are operators charged under the $\mathbb{Z}_{L}$ subgroup of the flavor group $SU(N_{f})$. We then have:
\begin{equation}\label{ineqaqed}
\min\{E_{\mathrm{shift}}, ~E_{\mathbb{Z}_{L}}\} \lsim E_{\mathrm{magnetic}}~.
\end{equation}

\subsubsection{Adjoint Higgsing Example}

As with the inequalities derived for axion-QCD, we expect that in examples \eqref{ineqamax} and \eqref{ineqaqed} are enforced automatically by self consistency.  Let us illustrate this in a concrete flow ending in axion-Maxwell theory.

In the UV we consider a model of $SU(2)$ axion-Yang-Mills coupled to a real adjoint Higgs field $\Phi$, and potential $V(\Phi)$.  The action is:
\begin{equation}
\frac{1}{2}\int da \wedge *da +\frac{1}{g^{2}}\int \mathrm{Tr}(F\wedge *F)-\frac{i}{8\pi^{2}f}\int a \mathrm{Tr}(F\wedge F)+ \int \mathrm{Tr}\left(D\Phi \wedge *D\Phi\right)-V(\Phi)~.
\end{equation}
Note in particular that we have set the integral coupling constant $K=1$ above.  This means that this UV action has symmetry $\mathbb{Z}_{2}^{(1)}$ (arising from the center of the gauge group), and $U(1)^{(2)}$, but no discrete shift symmetry of the axion.
 
We now consider a potential such that $\langle \mathrm{Tr}(\Phi^{2})\rangle=v^{2}$ thus higgsing the gauge group from $SU(2)$ to $U(1)$.  Importantly, this means that the minimum instanton number is multiplied by two: $\mathrm{Tr}(F\wedge F)\rightarrow 2 F\wedge F$, where the right-hand-side is the remaining abelian gauge field.  Therefore at low-energies we find axion-Maxwell theory (plus an irrelevant decoupled scalar descending from $\Phi$) with discrete coupling constant $K=2$.  Therefore, at low-energies we arrive in a theory with non-trivial higher group structure extending the emergent $U(1)^{(1)}_{m}$ magnetic one-form symmetry.

It is straightforward to determine the energy scales associated with the emergent symmetries:
\begin{itemize}
\item The $\mathbb{Z}_{2}^{(1)}$ electric symmetry is present in both the UV and IR.  Thus we have $E_{\mathrm{screen}}\rightarrow \infty$.
\item The $U(1)^{(1)}_{m}$ magnetic one-form symmetry emerges at the Higgsing scale, so $E_{\mathrm{magnetic}}\approx v$.  One way to understand this is that, in terms of UV non-abelian fields, the long-distance gauge field is $\frac{1}{v}\mathrm{Tr}(\Phi F)$.  Therefore, the scale $v$ is where its correlation functions begin to differ from those of an Abelian gauge field.

\item The $\mathbb{Z}_{2}^{(0)}$ emergent zero form symmetry is broken in the UV by instantons that generate a potential for the axion.  One can estimate this potential using an instanton gas approximation 
\begin{equation}
V(a)\approx g^{4\alpha}v^{4}\exp\left(-\frac{8\pi^{2}}{g^{2}}\right)\cos(a/f)~.
\end{equation}
Here, $g$ is the coupling evaluated at the scale $v$ of Higgsing, and $\alpha$ is a constant fixed by the $\beta$ function of the theory.  This formula arises by estimating the integral over instanton moduli space where the size modulus of the instanton is cut off at the scale $gv$ where the W-bosons become massive.  The energy cost of a shift $a\rightarrow a+\pi f$, then sets the scale $E_{\mathrm{shift}}$:
\begin{equation}
E_{\mathrm{shift}}\approx g^{\alpha}v \ \exp\left(-\frac{2\pi^{2}}{g^{2}}\right)~.
\end{equation}
\end{itemize}

We can now apply the inequality \eqref{ineqamax}:
\begin{equation}
E_{\mathrm{shift}}\lsim E_{\mathrm{magnetic}} \Longleftrightarrow g^{\alpha} \ \exp\left(-\frac{2\pi^{2}}{g^{2}}\right) \lsim 1~.
\end{equation}
Thus, the inequality is simply the statement that at the higgsing scale $v$, where $g$ is evaluated, the theory is weakly coupled $g(v)<<1$.  Of course, if this is not true, the theory confines before higgsing and our analysis leading to axion-maxwell theory in the IR breaks down.  Therefore \eqref{ineqamax} is automatic as expected.

\section*{Acknowledgements}
We thank T. Dumitrescu, Z. Komargodski, M. Reece, and L.T. Wang for discussions and comments. TDB is supported by the Mafalda and Reinhard Oehme Postdoctoral Fellowship in the Enrico Fermi Institute at the University of Chicago and in part by  the US Department of Energy DE-SC0009924.  The work of CC is supported by the US Department of Energy DE-SC0021432 and by the Simons Collaboration on Global Categorical Symmetries.
\bibliographystyle{utphys}
\bibliography{3GroupBib}

\providecommand{\href}[2]{#2}\begingroup\raggedright\begin{thebibliography}{10}

\bibitem{Kim:2008hd}
J.~E. Kim and G.~Carosi, ``{Axions and the Strong CP Problem},''
  \href{http://dx.doi.org/10.1103/RevModPhys.82.557}{{\em Rev. Mod. Phys.}
  {\bfseries 82} (2010) 557--602},
  \href{http://arxiv.org/abs/0807.3125}{{\ttfamily arXiv:0807.3125 [hep-ph]}}.
  [Erratum: Rev.Mod.Phys. 91, 049902 (2019)].

\bibitem{Marsh:2015xka}
D.~J.~E. Marsh, ``{Axion Cosmology},''
  \href{http://dx.doi.org/10.1016/j.physrep.2016.06.005}{{\em Phys. Rept.}
  {\bfseries 643} (2016) 1--79},
  \href{http://arxiv.org/abs/1510.07633}{{\ttfamily arXiv:1510.07633
  [astro-ph.CO]}}.

\bibitem{Svrcek:2006yi}
P.~Svrcek and E.~Witten, ``{Axions In String Theory},''
  \href{http://dx.doi.org/10.1088/1126-6708/2006/06/051}{{\em JHEP} {\bfseries
  06} (2006) 051}, \href{http://arxiv.org/abs/hep-th/0605206}{{\ttfamily
  arXiv:hep-th/0605206}}.

\bibitem{Gaiotto:2014kfa}
D.~Gaiotto, A.~Kapustin, N.~Seiberg, and B.~Willett, ``{Generalized Global
  Symmetries},'' \href{http://dx.doi.org/10.1007/JHEP02(2015)172}{{\em JHEP}
  {\bfseries 02} (2015) 172}, \href{http://arxiv.org/abs/1412.5148}{{\ttfamily
  arXiv:1412.5148 [hep-th]}}.

\bibitem{Lake:2018dqm}
E.~Lake, ``{Higher-form symmetries and spontaneous symmetry breaking},''
  \href{http://arxiv.org/abs/1802.07747}{{\ttfamily arXiv:1802.07747
  [hep-th]}}.

\bibitem{Coleman:1973ci}
S.~R. Coleman, ``{There are no Goldstone bosons in two-dimensions},''
  \href{http://dx.doi.org/10.1007/BF01646487}{{\em Commun. Math. Phys.}
  {\bfseries 31} (1973) 259--264}.

\bibitem{Green:1984sg}
M.~B. Green and J.~H. Schwarz, ``{Anomaly Cancellation in Supersymmetric D=10
  Gauge Theory and Superstring Theory},''
  \href{http://dx.doi.org/10.1016/0370-2693(84)91565-X}{{\em Phys. Lett. B}
  {\bfseries 149} (1984) 117--122}.

\bibitem{Kapustin:2013uxa}
A.~Kapustin and R.~Thorngren, ``{Higher symmetry and gapped phases of gauge
  theories},'' \href{http://arxiv.org/abs/1309.4721}{{\ttfamily arXiv:1309.4721
  [hep-th]}}.

\bibitem{Kapustin:2014zva}
A.~Kapustin and R.~Thorngren, ``{Anomalies of discrete symmetries in various
  dimensions and group cohomology},''
  \href{http://arxiv.org/abs/1404.3230}{{\ttfamily arXiv:1404.3230 [hep-th]}}.

\bibitem{Tachikawa:2017gyf}
Y.~Tachikawa, ``{On gauging finite subgroups},''
  \href{http://dx.doi.org/10.21468/SciPostPhys.8.1.015}{{\em SciPost Phys.}
  {\bfseries 8} no.~1, (2020) 015},
  \href{http://arxiv.org/abs/1712.09542}{{\ttfamily arXiv:1712.09542
  [hep-th]}}.

\bibitem{Cordova:2018cvg}
C.~C\'ordova, T.~T. Dumitrescu, and K.~Intriligator, ``{Exploring 2-Group
  Global Symmetries},'' \href{http://dx.doi.org/10.1007/JHEP02(2019)184}{{\em
  JHEP} {\bfseries 02} (2019) 184},
  \href{http://arxiv.org/abs/1802.04790}{{\ttfamily arXiv:1802.04790
  [hep-th]}}.

\bibitem{Benini:2018reh}
F.~Benini, C.~C\'ordova, and P.-S. Hsin, ``{On 2-Group Global Symmetries and
  their Anomalies},'' \href{http://dx.doi.org/10.1007/JHEP03(2019)118}{{\em
  JHEP} {\bfseries 03} (2019) 118},
  \href{http://arxiv.org/abs/1803.09336}{{\ttfamily arXiv:1803.09336
  [hep-th]}}.

\bibitem{Seiberg:2018ntt}
N.~Seiberg, Y.~Tachikawa, and K.~Yonekura, ``{Anomalies of Duality Groups and
  Extended Conformal Manifolds},''
  \href{http://dx.doi.org/10.1093/ptep/pty069}{{\em PTEP} {\bfseries 2018}
  no.~7, (2018) 073B04}, \href{http://arxiv.org/abs/1803.07366}{{\ttfamily
  arXiv:1803.07366 [hep-th]}}.

\bibitem{Cordova:2019uob}
C.~C\'ordova, D.~S. Freed, H.~T. Lam, and N.~Seiberg, ``{Anomalies in the Space
  of Coupling Constants and Their Dynamical Applications II},''
  \href{http://dx.doi.org/10.21468/SciPostPhys.8.1.002}{{\em SciPost Phys.}
  {\bfseries 8} no.~1, (2020) 002},
  \href{http://arxiv.org/abs/1905.13361}{{\ttfamily arXiv:1905.13361
  [hep-th]}}.

\bibitem{Hidaka:2020iaz}
Y.~Hidaka, M.~Nitta, and R.~Yokokura, ``{Higher-form symmetries and 3-group in
  axion electrodynamics},''
  \href{http://dx.doi.org/10.1016/j.physletb.2020.135672}{{\em Phys. Lett. B}
  {\bfseries 808} (2020) 135672},
  \href{http://arxiv.org/abs/2006.12532}{{\ttfamily arXiv:2006.12532
  [hep-th]}}.

\bibitem{Hidaka:2020izy}
Y.~Hidaka, M.~Nitta, and R.~Yokokura, ``{Global 3-group symmetry and 't Hooft
  anomalies in axion electrodynamics},''
  \href{http://arxiv.org/abs/2009.14368}{{\ttfamily arXiv:2009.14368
  [hep-th]}}.

\bibitem{Cordova:2020tij}
C.~C\'{o}rdova, T.~T. Dumitrescu, and K.~Intriligator, ``{2-Group Global
  Symmetries and Anomalies in Six-Dimensional Quantum Field Theories},''
  \href{http://arxiv.org/abs/2009.00138}{{\ttfamily arXiv:2009.00138
  [hep-th]}}.

\bibitem{Kim:1979if}
J.~E. Kim, ``{Weak Interaction Singlet and Strong CP Invariance},''
  \href{http://dx.doi.org/10.1103/PhysRevLett.43.103}{{\em Phys. Rev. Lett.}
  {\bfseries 43} (1979) 103}.

\bibitem{Shifman:1979if}
M.~A. Shifman, A.~Vainshtein, and V.~I. Zakharov, ``{Can Confinement Ensure
  Natural CP Invariance of Strong Interactions?},''
  \href{http://dx.doi.org/10.1016/0550-3213(80)90209-6}{{\em Nucl. Phys. B}
  {\bfseries 166} (1980) 493--506}.

\bibitem{Peccei:1977hh}
R.~Peccei and H.~R. Quinn, ``{CP Conservation in the Presence of Instantons},''
  \href{http://dx.doi.org/10.1103/PhysRevLett.38.1440}{{\em Phys. Rev. Lett.}
  {\bfseries 38} (1977) 1440--1443}.

\bibitem{Peccei:1977ur}
R.~Peccei and H.~R. Quinn, ``{Constraints Imposed by CP Conservation in the
  Presence of Instantons},''
  \href{http://dx.doi.org/10.1103/PhysRevD.16.1791}{{\em Phys. Rev. D}
  {\bfseries 16} (1977) 1791--1797}.

\bibitem{Cordova:2019jnf}
C.~C\'ordova, D.~S. Freed, H.~T. Lam, and N.~Seiberg, ``{Anomalies in the Space
  of Coupling Constants and Their Dynamical Applications I},''
  \href{http://dx.doi.org/10.21468/SciPostPhys.8.1.001}{{\em SciPost Phys.}
  {\bfseries 8} no.~1, (2020) 001},
  \href{http://arxiv.org/abs/1905.09315}{{\ttfamily arXiv:1905.09315
  [hep-th]}}.

\bibitem{Cordova:2019bsd}
C.~C\'ordova and K.~Ohmori, ``{Anomaly Obstructions to Symmetry Preserving
  Gapped Phases},'' \href{http://arxiv.org/abs/1910.04962}{{\ttfamily
  arXiv:1910.04962 [hep-th]}}.

\bibitem{Cordova:2017kue}
C.~C\'ordova, P.-S. Hsin, and N.~Seiberg, ``{Time-Reversal Symmetry, Anomalies,
  and Dualities in (2+1)$d$},''
  \href{http://dx.doi.org/10.21468/SciPostPhys.5.1.006}{{\em SciPost Phys.}
  {\bfseries 5} no.~1, (2018) 006},
  \href{http://arxiv.org/abs/1712.08639}{{\ttfamily arXiv:1712.08639
  [cond-mat.str-el]}}.

\bibitem{Gaiotto:2017yup}
D.~Gaiotto, A.~Kapustin, Z.~Komargodski, and N.~Seiberg, ``{Theta, Time
  Reversal, and Temperature},''
  \href{http://dx.doi.org/10.1007/JHEP05(2017)091}{{\em JHEP} {\bfseries 05}
  (2017) 091}, \href{http://arxiv.org/abs/1703.00501}{{\ttfamily
  arXiv:1703.00501 [hep-th]}}.

\end{thebibliography}\endgroup

\end{document}